\documentclass[nofootinbib]{elsart}

\usepackage{graphicx}
\usepackage{dcolumn}
\usepackage{bm}
\usepackage{xfrac}
\usepackage[ulem=normalem]{changes}
\begin{document}

\begin{frontmatter}
\title{Gamow--Teller transitions and neutron--proton-pair transfer reactions}
\author[ganil]{P.~Van~Isacker}
and
\author[berkeley]{A.~O.~Macchiavelli}

\address[ganil]{Grand Acc\'el\'erateur National d'Ions Lourds, CEA/DRF--CNRS/IN2P3,\\
Bvd Henri Becquerel, F-14076 Caen, France}
\address[berkeley]{Nuclear Science Division, Lawrence Berkeley National Laboratory\\
Berkeley, California 94720, USA}

\begin{abstract}
We propose a schematic model of nucleons moving in spin--orbit partner levels, $j=l\pm{\sfrac12}$,
to explain Gamow--Teller and two-nucleon transfer data in $N=Z$ nuclei above $^{40}$Ca.
Use of the $LS$ coupling scheme provides a more transparent approach
to interpret the structure and reaction data.
We apply the model to the analysis of charge-exchange,
$^{42}$Ca($^3$He,t)$^{42}$Sc, and np-transfer, $^{40}$Ca($^3$He,p)$^{42}$Sc, reactions data
to define the elementary modes of excitation 
in terms of both isovector and isoscalar pairs,
whose properties can be determined by adjusting the parameters of the model
(spin--orbit splitting, isovector pairing strength and quadrupole  matrix element) to the available data.
The overall agreement with experiment suggests
that the approach captures the main physics ingredients
and provides the basis for a boson approximation that can be extended to heavier nuclei. 
Our analysis also reveals that the SU(4)-symmetry limit is not realized in $^{42}$Sc.
\end{abstract}
\end{frontmatter}

In two recent papers Fujita {\it et al.}~\cite{Fujita14,Fujita15}
report on results of ($^3$He,t) charge-exchange experiments
that determine Gamow--Teller (GT) strength
in nuclei with mass numbers $A=42$, 46, 50 and 54.
They observe a concentration of most of the GT strength in the lowest $1^+$ state at 0.611~MeV
in the $^{42}{\rm Ca}\rightarrow {^{42}{\rm Sc}}$ reaction
and, as $A$ increases, a migration of this strength to higher energies.
Both features can be reproduced
either by a shell-model calculation with a realistic interaction in the $pf$ shell
or by calculations in the quasi-particle random phase approximation
that include an isoscalar (or spin-triplet) interaction.
The migration of the strength towards higher energies with $A$
can be understood intuitively as a result of the increasing importance
of the $\nu0f_{7/2}\rightarrow\pi0f_{5/2}$ component of the GT transition.
The low-energy strength in $^{42}$Sc is more difficult to fathom
and is attributed to the isoscalar component
of the residual nuclear interaction.
As a result, the authors~\cite{Fujita14,Fujita15} claim
the $1^+_1$ level in $^{42}$Sc to be a ``low-energy super GT state'',
and its existence is attributed to the restoration of Wigner's SU(4) symmetry~\cite{Wigner37}.

Relevant to these studies are the results of earlier measurements of two-nucleon transfer 
using the $^{40}$Ca($^3$He,p)$^{42}$Sc reaction~\cite{Zurmuhle66,Puhlhofer68}.
The coherent properties of the transfer mechanism of the neutron--proton (np) pair,
in both isospins channels, provide a complementary tool
to probe the wave functions
of the low-lying $0^+$ and $1^+$  levels in $^{42}$Sc.
The comparable cross-sections to these states appear, {\sl a priori},
at odds with the ``super GT" conjecture above.

In this letter we propose an explanation of these observations,
assuming that the nucleons occupy two orbitals
with radial quantum number $n$, orbital angular momentum $l$
and total angular momentum $j=l\pm{\sfrac12}$.
This is the analogue of a single-$j$ approximation,
for example the $0f_{7/2}$ model~\cite{Cullen64},
but for an $l$ orbital.
Since the properties of the nuclear interaction are more transparent in $LS$ coupling,
we analyze the problem in this basis instead of the more usual $jj$ coupling.
Results are of course independent of the chosen basis
and generally intermediate between the two bases~\cite{Zeldes53}.
Two nucleons with isospin projection $T_z=\pm1$,
angular momentum $J=0$ and isospin $T=1$
have two possible components with $(LS)=(00)$ and $(11)$,
where $L$ refers to the orbital angular momentum of the two nucleons and $S$ to their spin.
Three states with $(JT)=(10)$ occur for a neutron-proton pair
and they are admixtures of $(LS)=(01)$, $(10)$ and $(21)$.
One state with $(JT)=(11)$ exists for $T_z=0$ and it has $(LS)=(11)$.
These are the only states that enter into the discussion
of the GT strength and np transfer.

The character of the eigenstates of a nuclear Hamiltonian in this basis
is first of all determined by the one-body spin--orbit term
\begin{equation}
\hat H_{\rm so}=
\epsilon_-\hat n_-+\epsilon_+\hat n_+
=\Delta\epsilon\,{\frac12}(\hat n_--\hat n_+)
+\bar\epsilon\,\hat n,
\label{e_fso}
\end{equation}
where $\hat n$ is the nucleon-number operator,
$\hat n_\pm$ are the nucleon-number operators for the two orbitals $j=l\pm{\sfrac12}$
with single-particle energies $\epsilon_\pm$,
$\Delta\epsilon\equiv\epsilon_--\epsilon_+$
and $\bar\epsilon\equiv{\frac12}(\epsilon_-+\epsilon_+)$.
The operator $\hat H_{\rm so}$
is a non-diagonal one-body operator in the $(LSJT)$ basis
with the following matrices:
\begin{itemize}
\item
for $(JT)=(01)$ in the basis $(LSJT)=(0001)$ and $(1101)$:
\begin{equation}
2\bar\epsilon+
\frac{\Delta\epsilon}{2l+1}
\left[\begin{array}{cc}
-1&\sqrt{4l(l+1)}\\[2ex]
\sqrt{4l(l+1)}&1
\end{array}\right];
\label{e_hmat01}
\end{equation}
\item
for $(JT)=(11)$ in the basis $(LSJT)=(1111)$:
\begin{equation}
2\bar\epsilon;
\label{e_hmat11}
\end{equation}
\item
for $(JT)=(10)$ in the basis $(LSJT)=(0110)$, $(1010)$ and $(2110)$:
\begin{equation}
2\bar\epsilon+
\frac{\Delta\epsilon}{2l+1}
\left[\begin{array}{ccc}
-1&-\displaystyle\sqrt{\frac{4l(l+1)}{3}}&0\\
-\displaystyle\sqrt{\frac{4l(l+1)}{3}}&-1&\displaystyle\sqrt{\frac{2(2l-1)(2l+3)}{3}}\\
0&\displaystyle\sqrt{\frac{2(2l-1)(2l+3)}{3}}&2
\end{array}\right].
\label{e_hmat10}
\end{equation}
\end{itemize}
For each $(JT)$ a complete set $(LSJT)$ is given
and therefore the diagonalization of the above matrices leads to the correct eigenvalues
$2\epsilon_-$, $\epsilon_-+\epsilon_+$ and/or $2\epsilon_+$.
Matrices for different $(JT)$ can be constructed likewise
but the ones given in Eqs.~(\ref{e_hmat01}) to (\ref{e_hmat10})
suffice for the applications considered below.

To $\hat H_{\rm so}$ must be added contributions from the two-body interaction $\hat V$,
which can have diagonal matrix elements $V_{LSJT}\equiv\langle LSJT|\hat V|LSJT\rangle$
as well as off-diagonal ones $\langle LSJT|\hat V|L'S'JT\rangle$,
where it is assumed that the interaction
is invariant under rotations in physical and isospin space
and therefore conserves $J$ and $T$.

The structure of the eigenstates
is mostly determined by the splitting $\Delta\epsilon$,
to which the interactions $V_{LSJT}$ provide a correction.
Off-diagonal matrix elements due to spin-dependent or tensor forces
are small compared to those induced by $\hat H_{\rm so}$
and can be neglected in this context,
$\langle LSJT|\hat V|L'S'JT\rangle\approx0$ if $(LS)\neq(L'S')$.
Furthermore, the nuclear interaction in spatially anti-symmetric states ($L$ odd) is weak,
$V_{11J1}\approx V_{1010}\approx0$.
These approximations follow from the short-range attractive nature
of the residual nuclear interaction
and are exactly satisfied by a delta interaction~\cite{Talmi93}.
They lead to a description of structural properties in terms of three essential quantities:
the spin--orbit splitting $\Delta\epsilon$,
and the isoscalar and isovector pairing strengths $V_{0110}$ and $V_{0001}$,
which we denote from now on as $g_0$ and $g_1$, respectively.
There is an additional dependence
on the quadrupole matrix element $V_{2110}$,
which appears in the $(JT)=(10)$ matrix,
but this dependence is weak
and the value of $V_{2110}$ can be estimated from data (see below).

To calculate various properties in the $LSJT$ basis,
we consider a general one-body operator with definite tensor character
$\lambda_l$ under ${\rm SO}_L(3)$,
$\lambda_s$ under ${\rm SO}_S(3)$,
coupled to total $\lambda_j$,
and $\lambda_t$ under ${\rm SO}_T(3)$.
It has the matrix elements
\begin{eqnarray}
&&\langle l^2LSJT|||
\sum_i[\hat t^{(\lambda_l)}_i\times\hat t^{(\lambda_s)}_i]^{(\lambda_j)}\hat t^{(\lambda_t)}_i
||| l^2L'S'J'T'\rangle
\nonumber\\&&\quad=
-2[\lambda_j][L][S][J][T][L'][S'][J'][T']
\textstyle{\langle l\|\hat t^{(\lambda_l)}\| l\rangle
\langle{\sfrac12}\|\hat t^{(\lambda_s)}\|{\sfrac12}\rangle
\langle{\sfrac12}\|\hat t^{(\lambda_t)}\|{\sfrac12}\rangle}
\nonumber\\&&\quad\phantom{=}\times
(-)^{\lambda_l+\lambda_s+\lambda_t}
\left\{\!\!\begin{array}{ccc}
L&S&J\\L'&S'&J'\\\lambda_l&\lambda_s&\lambda_j
\end{array}\!\!\right\}
\Biggl\{\!\!\begin{array}{ccc}
L&L'&\lambda_l\\l&l&l
\end{array}\!\!\Biggr\}
\Biggl\{\!\!\begin{array}{ccc}
S&S'&\lambda_s\\{\sfrac12}&{\sfrac12}&{\sfrac12}
\end{array}\!\!\Biggr\}
\Biggl\{\!\!\begin{array}{ccc}
T&T'&\lambda_t\\{\sfrac12}&{\sfrac12}&{\sfrac12}
\end{array}\!\!\Biggr\},
\label{e_me}
\end{eqnarray}
where the symbols in curly brackets are $6j$- and $9j$-coefficients~\cite{Talmi93}
and with $[x]\equiv\sqrt{2x+1}$.
The triple bars on the left-hand side
indicate that the matrix element is reduced in $J$ and $T$
while the double-barred matrix elements on the right-hand side
are singly reduced in $L$, $S$ or $T$.
With this expression one can calculate
matrix elements of the M1 operator ($\lambda_j=1$),
which has spin $(\lambda_l,\lambda_s)=(0,1)$,
orbital $(\lambda_l,\lambda_s)=(1,0)$
and tensor $(\lambda_l,\lambda_s)=(1,1)$ parts
of both isoscalar ($\lambda_t=0$) and isovector ($\lambda_t=1$) character.
For the GT operator one takes $(\lambda_l,\lambda_s,\lambda_j,\lambda_t)=(0,1,1,1)$.
One finds three allowed GT transitions,
namely $(LS)=(00)\rightarrow(01)$, $(11)\rightarrow(10)$ and $(11)\rightarrow(11)$.
The strengths are independent of the orbital angular momentum $l$
with $JT$-reduced matrix elements given by $-\sqrt{18}$, $\sqrt{6}$ and $-\sqrt{24}$, respectively.

To obtain predictions for the np-transfer strengths,
we treat the ground state of $^{40}$Ca as the vacuum $|{\rm o}\rangle$
and write the wave functions of the $0^+_i$ and $1^+_i$ states in $^{42}$Sc as
\begin{eqnarray}
|{^{42}{\rm Sc}}(0^+_i)\rangle&=&
\alpha^i_{00}|l^20001\rangle+\alpha^i_{11}|l^21101\rangle,
\nonumber\\
|{^{42}{\rm Sc}}(1^+_i)\rangle&=&
\alpha^i_{01}|l^20110\rangle+\alpha^i_{10}|l^21010\rangle+\alpha^i_{21}|l^22110\rangle,
\label{e_tf1}
\end{eqnarray}
with coefficients $\alpha^i_{LS}$ obtained from the diagonalization
of the matrices~(\ref{e_hmat01}) and~(\ref{e_hmat10}).
In $LS$ coupling the $L=0$ transfer strengths follow naturally from
\begin{eqnarray}
|\langle{^{42}{\rm Sc}}(0^+_i)||A^\dag_{L=0,S=J=0,T=1}||{^{40}{\rm Ca}}(0^+_1)\rangle|^2&=&(\alpha^i_{00})^2,
\nonumber\\
|\langle{^{42}{\rm Sc}}(1^+_i)||A^\dag_{L=0,S=J=1,T=0}||{^{40}{\rm Ca}}(0^+_1)\rangle|^2&=&3(\alpha^i_{01})^2,
\label{e_tf2}
\end{eqnarray}
where $A^\dag_{LSJT}$ is a two-nucleon creation operator.

We apply the above schematic model to the properties of $A=42$ nuclei.
We fix the spin--orbit splitting to its value taken in Refs.~\cite{Fujita14,Fujita15},
$\Delta\epsilon=6$~MeV,
and vary the pairing strengths $g_0$ and $g_1$.
We take as a first estimate equal isoscalar and isovector pairing strengths,
and allow for a variation of 15~\% of the isoscalar strength,
that is, we consider $g_0=g_1/x$ with $x$ between 0.85 and 1.15,
indicated by shaded bands around the `canonical' estimate $g_0=g_1$.
In this way the sensitivity of the various properties
to the ratio of isoscalar-to-isovector pairing strengths is highlighted.
The quadrupole matrix element $V_{2110}\approx V_{2021}$ is fixed
such that the excitation energy of the $2^+_1$ level in $^{42}$Ca (1.525~MeV) is reproduced.
Essentially the same results are obtained
if $V_{2110}$ is varied within a wide range.

\begin{figure}
\centering
\includegraphics[height=9cm]{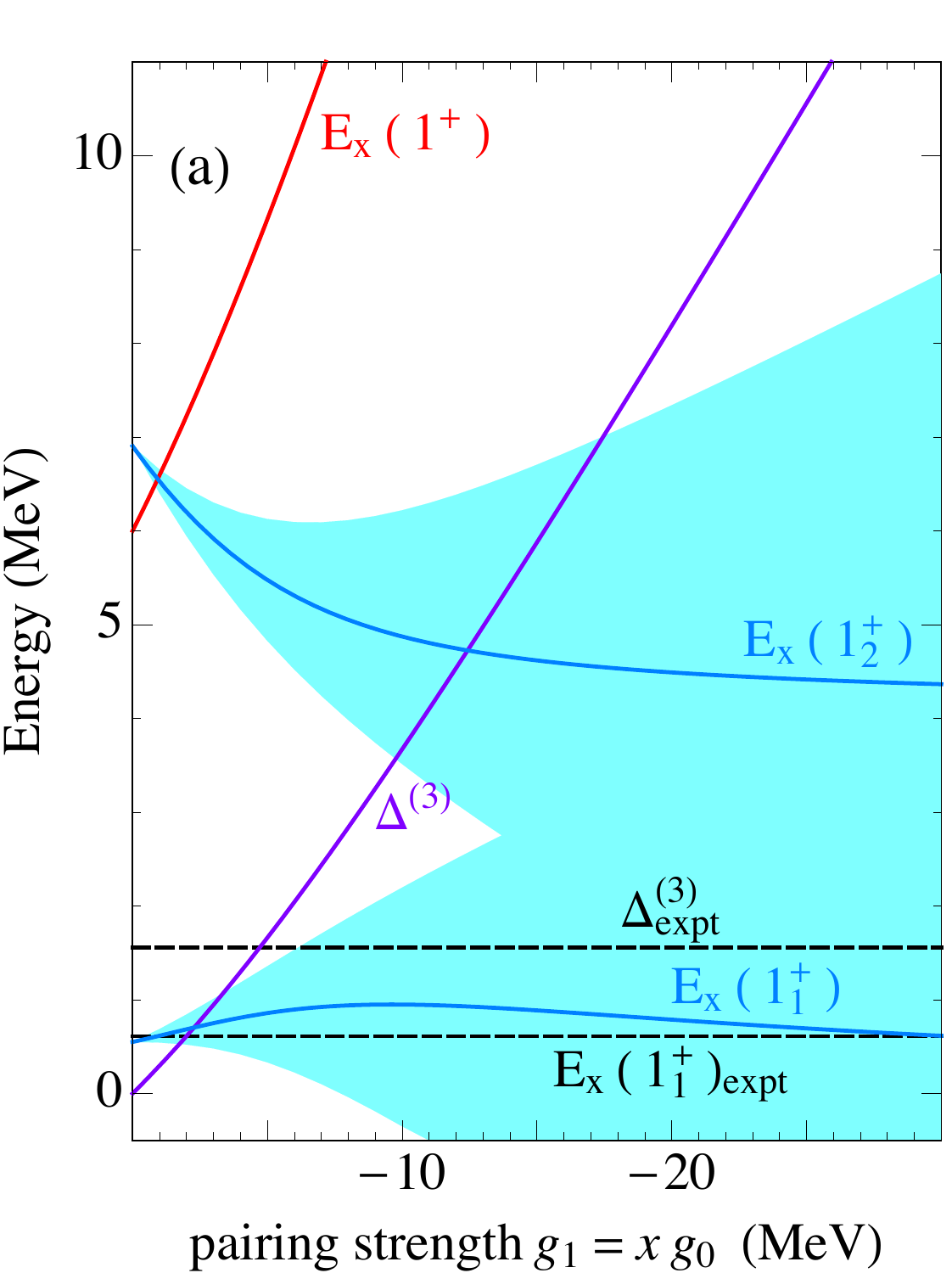}
\includegraphics[height=9cm]{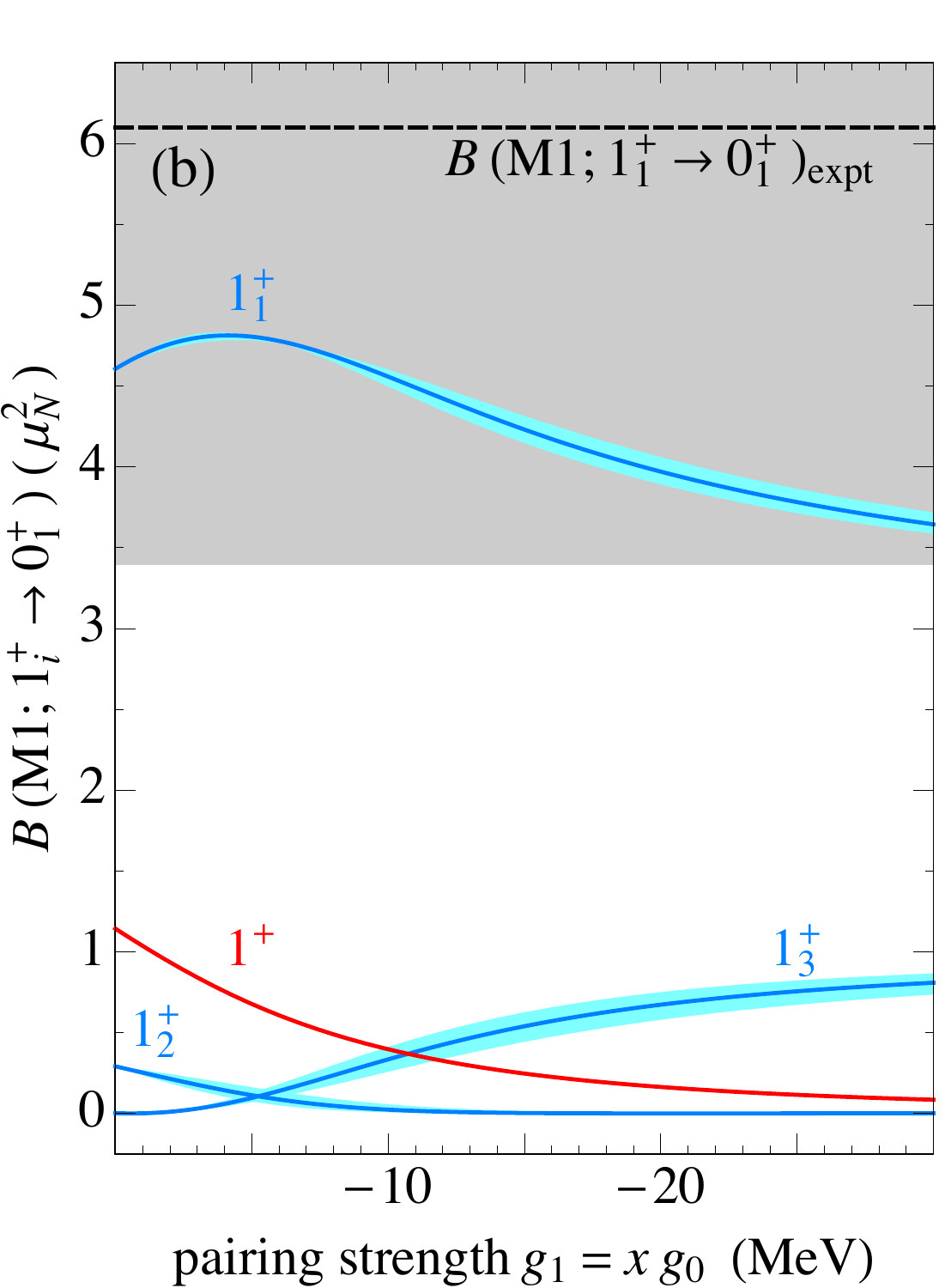}\\
\includegraphics[height=9cm]{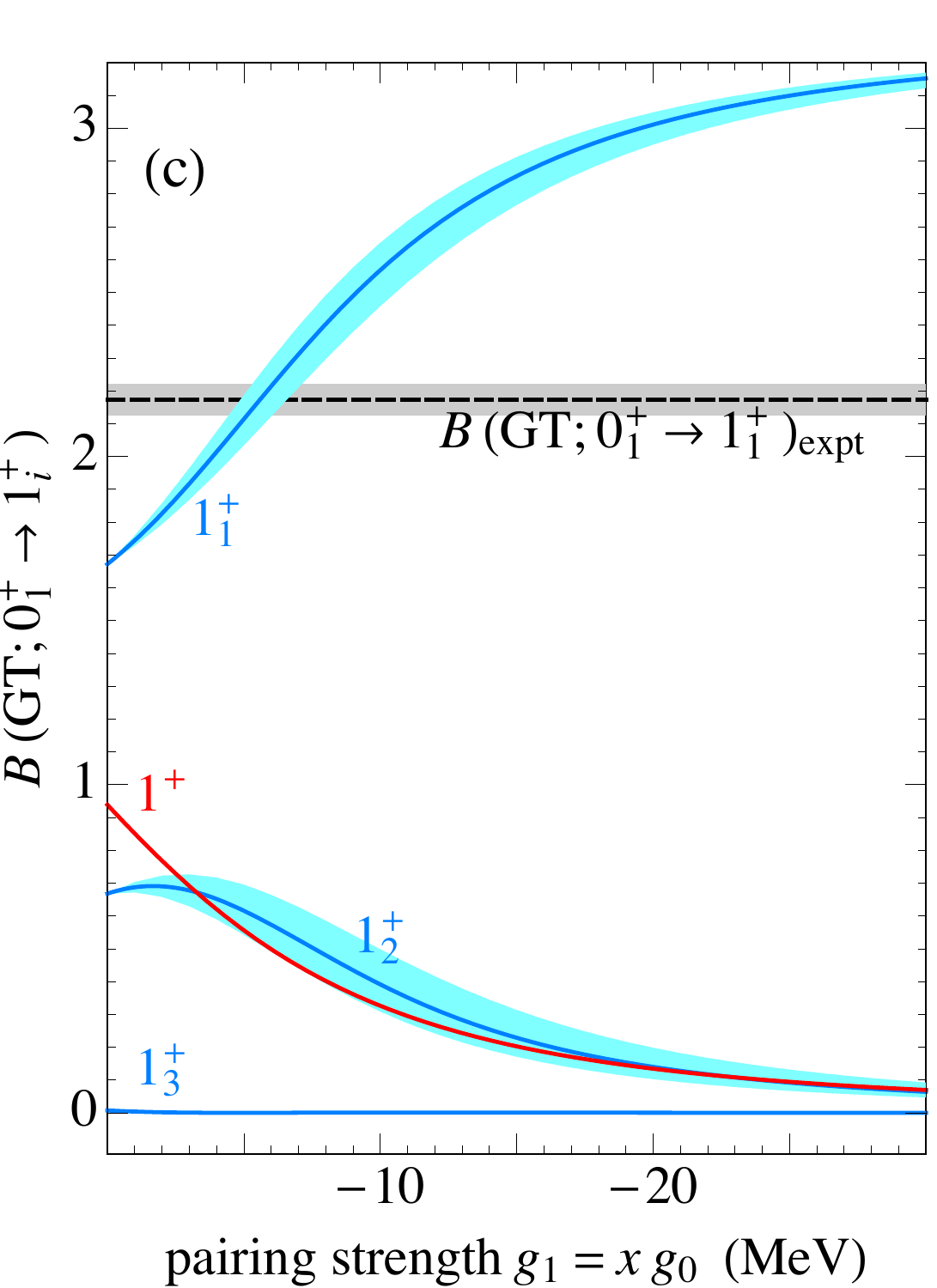}
\includegraphics[height=9cm]{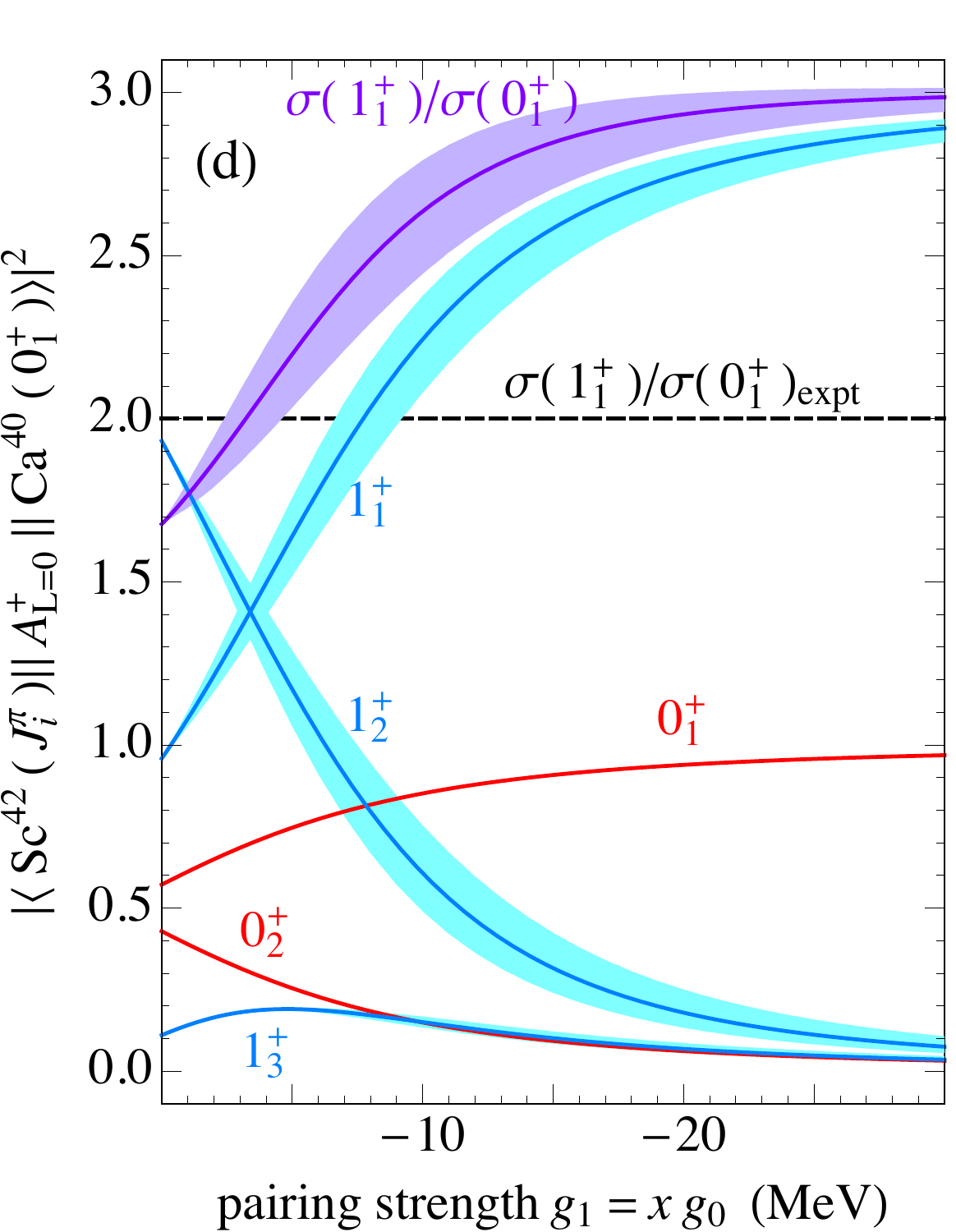}
\caption{\label{f_results}  
(Color  online)
Results for
(a) excitation energies in $^{42}$Sc and the pairing gap $\Delta^{(3)}$,
(b) $B({\rm M1};1^+_i\rightarrow0^+_1)$ values in $^{42}$Sc,
(c) $B({\rm GT};0^+_1\rightarrow1^+_i)$ strength
in the $^{42}{\rm Ca}\rightarrow {^{42}{\rm Sc}}$ reaction
and (d) cross-sections for the $^{40}{\rm Ca}\rightarrow {^{42}{\rm Sc}}$ reaction.
Experimental values are indicated by the black dashed lines with uncertainties in gray.
The results of the schematic model
are in blue for $T=0$ and in red for $T=1$ states.
The calculated pairing gap $\Delta^{(3)}$ in panel (a)
and the calculated ratio of cross-sections $\sigma(1^+_1)/\sigma(0^+_1)$ in panel (d) are in purple.
All curves are for equal isoscalar and isovector strengths $g_0=g_1$,
and the shaded areas around them are obtained
for $g_0=g_1/x$ with $x$ varying between 0.85 and 1.15.}
\end{figure}
Results are summarized in Fig.~\ref{f_results}.
Panel (a) shows the excitation energies of levels in $^{42}$Sc
(relative to the $0^+$ level)
as a function of the pairing strengths.
The $1^+_1$ level is at an essentially constant energy for $g_0=g_1$
but its energy is very sensitive to the ratio of the two pairing strengths.
The near-degeneracy of the $(JT)=(01)$ and $(10)$ states, therefore,
cannot be used as an indication of Wigner's SU(4) symmetry in $^{42}$Sc,
which is only realized in the extreme limit of $g_0=g_1\gg\Delta\epsilon$.
On the other hand, it is to be expected that the isovector pairing strength $g_1$
can be constrained from the corresponding observed pairing gap.
The ``pairing gap'' for the odd-mass nucleus $^{41}$Ca,
that is, the binding-energy difference
\begin{equation}
\Delta^{(3)}={\frac12}\left[{\rm BE}({^{42}{\rm Ca}})+{\rm BE}({^{40}{\rm Ca}})-2{\rm BE}({^{41}{\rm Ca}})\right],
\label{e_gap}
\end{equation}
is the only quantity of this kind that is available within our schematic model.
Its experimental value of 1.5585~MeV is also shown in panel (a) of Fig.~\ref{f_results}
and fixes the isovector pairing strength to $g_1\approx-5$~MeV.

Panel (b) shows the $B({\rm M1};1^+_i\rightarrow0^+_1)$ values in $^{42}$Sc,
calculated with a spin quenching factor of 0.74.
The M1 strength from the $1^+_1$ level is known experimentally~\cite{NNDC},
$B({\rm M1};1^+_1\rightarrow0^+_1)=6.1(2.7)$~$\mu_{\rm N}^2$,
but its error is too large to constrain the pairing strength.

Panel (c) shows the $B({\rm GT};0^+_1\rightarrow1^+_i)$ strength
for the $^{42}$Ca($^3$He,t)$^{42}$Sc charge-exchange reaction,
using the same quenching factor as in the spin part of the M1 operator.
Experimentally, most of the observed strength
is concentrated in the $1^+$ state at 0.611~MeV~\cite{Fujita14,Fujita15},
which is indicated in the figure,
with some fragmentation at higher energies (not shown).
This is in qualitative agreement with the schematic model,
where the main strength is indeed found in the $1^+_1$ level
with some minor components in two excited $1^+$ states with $T=0$ and $T=1$, respectively.
Note also that the uncertainty associated with the ratio of pairing strengths
is fairly small for the GT strength.

Finally, panel (d) shows results for the $^{40}{\rm Ca}\rightarrow {^{42}{\rm Sc}}$ np-transfer reaction.
We pay particular attention to the ratio of squared amplitudes from Eq.~(\ref{e_tf2})
that gives directly the ratio of cross-sections, $\sigma(1^+_1)/\sigma(0^+_1)$.
The experimental ratio, $\sim$2, is consistent with our results
at the value of the pairing strength $g_1$,
derived from the $B$(GT) measurement and from the pairing gap $\Delta^{(3)}$.
A better agreement would be expected
by introducing small admixtures of the $p_{3/2}$ and $p_{1/2}$ orbits,
not included in the schematic model.
We note here that we have also checked the calculated ratio in the $jj$-coupling scheme
by transforming the pair amplitudes
into their corresponding $f_{7/2}^2$, $f_{5/2}^2$ and $f_{7/2}f_{5/2}$ (for $T=0$) components,
which were used in DWUCK4~\cite{Kunz} 
to calculate the cross-sections at forward angles ($L=0$ transfer).

\begin{table}
\caption{Summary of available experimental data~\cite{Fujita15,Puhlhofer68,NNDC}
and the results of the schematic model for the adopted values of the parameters.}
\label{t_summary}
\renewcommand{\arraystretch}{1.1}
\begin{tabular}{c|ccc|ccc}
\hline\hline
Probe &\multicolumn{2}{c}{$^{42}$Sc level} &&Experiment &Schematic\\
           &$J^\pi$ &$E_{\rm x}$~(MeV) &&&model$^a$\\
\hline
($^3$He,p) &&&&\multicolumn{2}{c}{Relative Intensity}\\
&$0^+_1$&0 &&1&1 \\
&$1^+_1$&0.61 &&2 &2.20(17)\\           
&$(0^+,1^+)$&1.89 &&0.17 &---\\
&$1^+$&3.69 &&1.3 &1.57(16)\\        
&$1^+$&3.86 &&0.38 &---\\        
\hline                     
($^3$He,t) &&&&\multicolumn{2}{c}{$B$(GT)}\\
&$1^+_1$&0.61 &&2.17(5) &2.11(8)\\
&$1^+$&1.89 &&0.097(3) &---\\
&$1^+$&3.69 &&0.127(3) &0.62(8)\\
\hline                     
Lifetime &&&&\multicolumn{2}{c}{$B$(M1)~($\mu^2_{\rm N}$)}\\
DSAM &$1^+_1$ &0.61 &&6.1(2.7) &4.80(2)\\
\hline\hline
\multicolumn{7}{l}{$^a$The theoretical uncertainties correspond to a $\pm15\%$ variation}\\
\multicolumn{7}{l}{\phantom{$^a$}in the isoscalar pairing strength $g_0$, as discussed in the text.}
\end{tabular}
\end{table}
The available experimental data~\cite{Fujita15,Puhlhofer68,NNDC}
are summarized in Table~\ref{t_summary}
and compared with the results of the schematic model.
The latter are obtained with
spin--orbit splitting $\Delta\epsilon=6$~MeV
and pairing strengths $g_0=g_1=-5$~MeV.
The isoscalar pairing strength $g_0$ is varied by 15\% around $g_1$
to obtain an estimate of the theoretical uncertainty.
Because of the restricted model space with only the $f$ orbital,
several observed levels are absent from the theory,
as indicated with a dashed line.

\begin{table}
\caption{$LS$-coupling amplitudes (in \%) of the yrast $0^+$ and $1^+$ states of $^{42}$Sc,
in the schematic model and for the KB3G interaction.}
\label{t_amplitudes}
\renewcommand{\arraystretch}{1.1}
\begin{tabular}{cc|ccccc|cccc}
\hline\hline
&&&\multicolumn{3}{c}{Schematic model}&&&\multicolumn{3}{c}{KB3G ($f^2$)}\\
\cline{4-6}\cline{9-11}
&&&$L=0$&$L=1$&$L=2$&&&$L=0$&$L=1$&$L=2$\\
\hline
$0^+_1$&&&75 &25 &---&&&73 &23 &---\\
$1^+_1$&&&55 &31 &14 &&&65 &26 &4\\           
\hline                     
\end{tabular}
\end{table}
We can now also study the components in $LS$ coupling of the yrast $0^+$ and $1^+$ states,
for which the schematic model should be reliable.
Table~\ref{t_amplitudes} lists the amplitudes of $0^+_1$ and $1^+_1$,
written in the $(LSJT)$ basis of Eqs.~(\ref{e_hmat01}) and~(\ref{e_hmat10}).
It is seen that the spatially unfavoured components ($L$ odd) are important,
which contradicts the assumption of SU(4) symmetry.
The fact that nevertheless a strong $B({\rm GT};0^+_1\rightarrow1^+_1)$ is observed
is due to the constructive addition
of the $L=0\rightarrow L=0$ and $L=1\rightarrow L=1$ transitions.
Also shown in Table~\ref{t_amplitudes} are the corresponding components
for the modified Kuo--Brown KB3G Hamiltonian~\cite{Poves01},
which is a realistic interaction for the entire $pf$ shell~\cite{Caurier05}.
The $f^2$ components carry the majority of the strength ($\sim$95\%)
and the mixing of spatially favoured and unfavoured components
is consistent with that found in the schematic model.

In summary,
charge-exchange and np-transfer reactions define the elementary modes
with isovector and isoscalar pairs
that are spatially favoured as well as unfavoured.
We have applied this approach 
to $^{40,42}{\rm Ca}\rightarrow {^{42}{\rm Sc}}$ reactions
to determine the nature of these elementary modes.
Good agreement with the experimental data suggests the adequacy of the model.
Although the $1^+_1$ state carries a large fraction of the GT strength, 
our analysis of both ($^3$He,t) and ($^3$He,p) reactions
points out that the SU(4)-symmetry limit is not reached,
as the spin--orbit potential breaks the $LS$-coupling scheme.
The elementary pairs thus determined can be treated as bosons,
leading to an interpretation of GT and np-transfer data in heavier $0f_{7/2}$ nuclei
in terms of an interacting boson model---an
approach which is currently under study~\cite{Isackerun}.
We believe that this may provide an intuitive and simple picture,
which is complementary to state-of-the-art shell-model calculations. 

\section*{Acknowledgements}
This material is based upon work supported in part
by the U.S. Department of Energy, Office of Science, Office of Nuclear Physics
under Contracts DE-FG02-10ER41700 (FUSTIPEN) and DE-AC02-05CH11231 (LBNL).


\end{document}